\documentclass[preprint,secnumarabic,amssymb, nobibnotes, aps, prd]{revtex4}

\usepackage{amssymb}
\usepackage{amsmath}
\newcommand{\tr}{\text{tr}}

\begin{document}
\title{Disentanglement by Dissipative Open System Dynamics}
\author{P.J.Dodd}
\email{peter.dodd@imperial.ac.uk}
\affiliation{Blackett Laboratory\\ Imperial College\\ London SW7 2BZ\\UK}
\date{10 December 2003}
\begin{abstract}
This paper investigates disentanglement as a result of evolution according to a class of master equations which include dissipation and interparticle interactions. Generalizing an earlier result of Di\'{o}si, the time taken for complete disentanglement is calculated (i.e. for disentanglement from any other system). The dynamics of two harmonically coupled oscillators is solved in order to study the competing effects of environmental noise and interparticle coupling on disentanglement. An argument
based on separability conditions for gaussian states is used to arrive at a set of conditions on the couplings sufficient for
all initial states to disentangle for good after a finite time.
\end{abstract}
\maketitle

%\begin{subequations}
%\begin{align}
%S = S_0 - S_\delta & = Z |_{\delta = 0} \notag \\
%& = Z_{0}
%\end{align}
%\end{subequations}

\section{Introduction}

Entanglement - the fact that to all density matrices can be written as

\begin{eqnarray}
\rho^{1\& 2}=\sum_i p_i \rho_i^1 \otimes \rho_i^2
\end{eqnarray}
with $p_i$ a probability distribution - has become one of the most notable differences between classical and
quantum mechanics. What at first sight seems a rather technical difference, lies underneath all `no local hidden-variables'
theorems which help set quantum mechanics so firmly apart from classical mechanics (Ref\cite{bell}, Ref\cite{wer}). More recently, entanglement has come to
be regarded as a resource on which all essentially quantum technologies are based (quantum computation, cryptography, e.t.c., see e.g. Ref\cite{chuang}).
An understanding of how entanglement comes to be destroyed by interaction with an environment is therefore important not only
to explaining how the classical world emerges from an underlying quantum one, but also to understanding and controlling
the effects of noise, which make quantum technologies such a practical challenge.

In Ref\cite{ruskai}, a
theorem was proved categorizing completely disentangling maps. We call a completely positive  trace preserving map

\begin{equation}
M:\mathcal{B}(\mathcal{H})\to\mathcal{B}(\mathcal{H})
\end{equation}
completely disentangling if

\begin{equation}
M\otimes 1_N: \mathcal{B}(\mathcal{H}\otimes\mathbb{C}^N)\to\mathcal{B}(\mathcal{H}\otimes\mathbb{C}^N)
\end{equation}
has $M\otimes 1_N(\rho^{1\& 2})$ disentangled for all density matrices
$\rho^{1\& 2}\in\mathcal{B}(\mathcal{H}\otimes\mathbb{C}^N)$ and for all positive integers $N$. Ruskai  Ref\cite{ruskai} showed that
$M$ is completely disentangling if, and only if

\begin{equation}
M(\rho)=\sum_k\tr(\mu_k\rho)m_k\label{dis}
\end{equation}
with $\{\mu_k\}$ a POVM and $m_k$ density matrices.

Di\'{o}si has taken advantage of this to show that certain markovian open-system evolutions result in a map of this form
after a certain finite time Ref\cite{diosi}. This finite time environment-induced effect at first seems quite strange when compared
with the more familiar asymptotic environment-induced decoherence effects of similar time-scale. But it should be noted
that separable density matrices form a convex, finite-volume region within the space of all density matrices. If the
equilibrium state is separable, we expect the evolution to `suck' states towards it along smooth trajectories which,
at a certain time, pass into the separable region around the equilibrium state. This picture makes clearer how disentanglement
occurs in finite time, whereas decoherence, where states are `sucked' smoothly towards a stable submanifold of measure zero
(the states diagonal in the relevant basis), can only be achieved asymptotically.

In Ref\cite{DK} Di\'{o}si and Kiefer showed that the wigner function evolution

\begin{equation}
\frac{\partial W}{\partial t}=-\frac{p}{m}\frac{\partial
W}{\partial q}  + D_{pp}\frac{\partial ^2
W}{\partial p^2}\label{Devol}
\end{equation}
results in a positive $P$ function after a finite time. Di\'{o}si used essentially the same technology to show 
that this evolution is also completely disentangling after a finite time. The positive $P$ function result itself implies
a disentanglement result if we consider the obvious extension of (\ref{Devol}) to an $N$-particle system: a
positive $P$ function is a classical probability distribution over the (separable) coherent-state projectors. The complete
disentanglement result if stronger however, and also gives disentanglement on the symmetric $N$-particle system.

This paper is an elaboration and continuation of Ref\cite{us}, and is designed to be read in conjunction with it.  In Ref\cite{us}, the relation between decoherence and disentanglement was discussed and a new technique, based on separability criteria for gaussian states, used to compute the finite-time disentanglement of general states in a bipartite system, when evolved according to (\ref{Devol}). The evolution of an E.P.R.-like state was solved and used to illustrate how the separability criteria come to be satisfied. In this paper we first focus on extending Di\'{o}si's argument regarding the complete disentanglement of a single system to more general, dissipative evolutions. In particular we will be interested
in evolutions which are of Lindblad form, i.e.

\begin{equation}
\frac{d\rho}{dt}= -i[H,\rho] + \sum_k \left(2L_k^\dagger\rho L_k- L_kL_k^\dagger\rho-\rho L_k^\dagger L_k\right)
\end{equation}
with non-zero dissipation. Then we will also go on to apply the technique of Ref\cite{us} to a harmonically coupled pair of particles, where there is competition between the coupling, which can generate entanglement, and the environmental influence, which tends to destroy it. We will concentrate on determining conditions which guarantee that the environment wins.

The layout of this paper is as follows. Below, in Section 2, we extend Di\'{o}si's complete disentanglement result, analyzing a more general and realistic evolution.  For a specific choice of Lindblad operator, we compute the complete disentanglement timescale for high temperatures relative to the dissipation. 
In Section 3, we then go on to consider the open system evolution of bipartite systems
where the particles are coupled. Complete disentanglement results cannot be immediately applied to this situation because
the interaction will tend to generate new entanglement between subsystems. We will use a different method, based on the
recent separability criteria for gaussian states,  which explicitly
produces a decomposition of the state into separable states.
In section 4, we conclude and make some remarks about the role of entanglement destruction in explaining classical behaviour.

\section{Di\'{o}si analysis for a general dissipative master eqn}

 We start with the following Markovian (quantum Fokker-Planck) master equation for the Wigner
function

\begin{equation}
\frac{\partial W}{\partial t}=-\frac{p}{m}\frac{\partial
W}{\partial q} + m\omega ^2 q\frac{\partial W}{\partial p} +
2\gamma \frac{\partial (pW)}{\partial p} + D_{qq}\frac{\partial ^2
W}{\partial q \partial q} + D_{pp}\frac{\partial ^2
W}{\partial p \partial p} + 2D_{qp}\frac{\partial ^2
W}{\partial q \partial p}\label{evol}
\end{equation}
This describes a harmonic oscillator interacting with an environment whose effects can be well approximated by white noise. In practice, this is true fairly generically of commonly encountered environments at high temperatures.

It can be shown (see e.g. Ref\cite{arnold}) that this master equation is of Lindblad form if, and only if,  the diffusion matrix $D$, defined by

\begin{equation}
D=\left( \begin{array}{cc}D_{qq} & D_{qp}\\
D_{qp} & D_{pp}\end{array}\right)
\end{equation}
 is positive and

\begin{equation}
\det D\geq \frac{\gamma^2}{4m^2}\label{D>0}
\end{equation}
In analysing this master equation, we will make much use of the Fourier transform

\begin{equation}
W= \int \frac{d^2 \bar{z}}{ (2\pi)^2} \bar{W}(\bar{z})e^{i\langle
z,\eta \bar{z}\rangle}
\end{equation}
where

\begin{equation}
\langle a, b\rangle=a^Tb
\end{equation}
is the standard euclidean inner product,

\begin{equation}
\eta=\left( \begin{array}{cc}0 & 1\\
-1 & 0\end{array}\right)
\end{equation}
and

\begin{equation}
\bar{z}=\left( \begin{array}{c}\bar{q} \\ \bar{p} \end{array}\right)
\end{equation}
and similarly without `bars'.

The Fourier transform $\bar{W}$  obeys the equation

\begin{equation}
\frac{\partial \bar{W}}{\partial t}=
\frac{\bar{p}}{m}\frac{\partial \bar{W}}{\partial \bar{q}}
-\bar{q}m\omega^2 \frac{\partial \bar{W}}{\partial \bar{p}} +
2\gamma \bar{q}\frac{\partial \bar{W}}{\partial \bar{q}}
- \bar{z}^T\bar{D}\bar{z}\bar{W}
\end{equation}
with

\begin{equation}
\bar{D}=\left( \begin{array}{cc}D_{pp} & -D_{qp}\\
-D_{qp} & D_{qq}\end{array}\right)
\end{equation}
which we can solve by characteristics.

\subsection{Specialization to the free case, $\omega=0$}

Di\'{o}si's argument is given in Ref\cite{us}. Here, we explain the argument as we go, for $\omega=0$, but considering $\gamma\neq 0$. The calculations for the case $\omega\neq 0$ are contained implicitly in the section below on  coupled systems. We do not pursue Di\'{o}si's analysis in detail for this case because, as noted below, oscillating terms destroy the monotonicity of disentanglement and therefore make an analytical treatment much harder. 

The gist of the argument is to show that, by smearing things out in phase-space, the propagator for the evolution quickly results in a positive $P$ function. This ability of open-system dynamics to smear out the negative parts of phase-space quasi-distributions has been studied before with regard to treating them as genuine probability distributions in pseudoclassical circumstances. Here, the crucial observation is that a map which always results in states with positive $P$ function is a completely disentangling map, of the form (\ref{dis}). All this is accessible because the propagators are gaussian in form, meaning the calculations can be carried through explicitly, and  because the properties of gaussian states are well understood.

We start with the evolution written in the form

\begin{equation}
\frac{\partial \bar{W}}{\partial t}=
\frac{\bar{p}}{m}\frac{\partial \bar{W}}{\partial \bar{q}}
+2\gamma \bar{q}\frac{\partial \bar{W}}{\partial \bar{q}}
- \bar{z}^T\bar{D}\bar{z}\bar{W}
\end{equation}
which is equivalent to

\begin{equation}
\frac{d \bar{W}(\bar{z}_t,t)}{d t}= -\bar{z}_t^T \bar{D}\bar{z}_t\bar{W}(\bar{z}_t,t)
\end{equation}
where the equations for the characteristics are 

\begin{eqnarray}
&&\frac{\text{d}\bar{q}_t}{\text{d}t}=-\frac{\bar{p}_t}{m}-2\gamma
\bar{q}_t \nonumber\\
&&\frac{\text{d}\bar{p}_t}{\text{d}t}=0
\end{eqnarray}
Since this set of equations is linear and Markovian we can write

\begin{equation}
\bar{z}_t=E_t \bar{z}
\end{equation}
where $E_t$ is a family of matrices affecting the evolution as a
1-parameter group.We easily find

\begin{equation}
E_t=\left( \begin{array}{cc}1 & -\left(\frac{1-e^{-2\gamma t}}{2\gamma m}\right)\\
0 & 1\end{array}\right)
\end{equation}
Setting

\begin{equation}
\mu _t =\int_0 ^t d\tau E_{\tau} ^T N E_{\tau}
\end{equation}
we then can solve as

\begin{eqnarray}
\bar{W}(\bar{z},t)&=& \bar{W}(E_t(E_t ^{-1}z),t) \nonumber\\
&=& \bar{W}(\bar{z}_{-t},0)\exp[-\langle \bar{z}_{-t},\mu_t
\bar{z}_{-t} \rangle]
\end{eqnarray}
Taking the inverse  Fourier transform, for the original Wigner function we have

\begin{eqnarray}
W(z,t)&=& \int \frac{d^2 \bar{z}}{(2\pi)^2}
\bar{W}(\bar{z}_{-t},0)\exp[-\langle \bar{z}_{-t},\mu_t
\bar{z}_{-t}\rangle] e^{i\langle z,\eta \bar{z}\rangle}
\nonumber\\&=&\int \frac{d^2 \bar{z}}{(2\pi)^2} \int d^2 z'
e^{-i\langle z',\eta \bar{z}_{-t}\rangle }W(z',0)e^{i\langle
z,\eta \bar{z}\rangle }\exp[-\langle \bar{z}_{-t},\mu_t
\bar{z}_{-t}\rangle] e^{i\langle z,\eta \bar{z}\rangle}\nonumber\\
&=&\int d^2 z' \int \frac{d^2 \bar{z}}{(2\pi)^2} |\det(E_t)|
\exp[i\langle \eta^{T}E_t ^{T}\eta z- z',\eta \bar{z}
\rangle]\exp[-\langle \bar{z},\mu_t \bar{z}\rangle]W_0 (z')
\end{eqnarray}
We will not be so interested in the evolution of the argument, nor the normalizing
prefactors, which take care of themselves. The middle Fourier
transform can be carried out to give the propagator as

\begin{eqnarray}
&&\int \frac{d^2 \bar{z}}{ (2\pi)^2}\exp i\langle z,\eta \bar{z}
\rangle \exp -\langle \bar{z},\mu_t \bar{z}
\rangle\nonumber\\&=&\int \frac{d^2 \bar{z}}{ (2\pi)^2} \exp
-\langle \bar{z} + i\frac{{\mu_t} ^{-1}}{2}\eta z, \mu_t(\bar{z} +
i\frac{{\mu _t }^{-1}}{2}\eta z)\rangle e^{-\frac{1}{4}\langle
\eta z,\mu_t ^{-1}\eta
z\rangle}\nonumber\\&=&\frac{1}{2\pi  \sqrt{\det \mu_t}}\exp{
-\frac{1}{4}\langle z,\eta^{T} {\mu_t} ^{-1} \eta z \rangle} \nonumber\\
&\sim & g(2\eta ^T \mu_t \eta)
\end{eqnarray}
The notation $g(\Sigma)$ denoting a normalised gaussian with covariance matrix $\Sigma$.
In short then

\begin{equation}
W_t(z)=g(2\eta ^T \mu_t \eta)\ast W_0( \eta^{T}E_t ^{T}\eta z)|\det E_t|
\end{equation}
$\ast$ meaning convolution.

We calculate $2\eta ^T \mu_t \eta$ to be given by

\begin{equation}
2\eta ^T \mu_t \eta=2\left( \begin{array}{cc}D_{pp}\left( \frac{4\gamma t + 4e^{-2\gamma t}-e^{-4\gamma t}-3}
{16\gamma ^3 m^2}\right)+2D_{qp}\frac{2\gamma t + e^{-2\gamma t}-1}{4\gamma ^2 m}+D_{qq}t
& D_{pp}\frac{2\gamma t + e^{-2\gamma t}-1}{4\gamma ^2 m}+D_{qq}t\\
D_{pp}\frac{2\gamma t + e^{-2\gamma t}-1}{4\gamma ^2 m}+D_{qq}t & D_{pp}t\end{array}\right)
\end{equation}
We will now need the fact that the $P$ function and Wigner function are related by

\begin{equation}
W(z)=g(C_{1/4})\ast P(z)
\end{equation}
with

\begin{equation}
C_{1/4}=\left( \begin{array}{cc}\frac{1}{\sqrt{2D_{pp}m}} & \frac{1}{2}\\
\frac{1}{2} & \sqrt{\frac{D_{pp}m}{2}}\end{array}\right)
\end{equation}
Because the convolution of gaussians factors, (where it makes sense) we have

\begin{equation}
P_t(z)=g(2\eta ^T \mu_t \eta-C_{1/4})\ast W_0( \eta^{T}E_t ^{T}\eta z)|\det E_t|
\end{equation}
and as for Wigner functions

\begin{equation}
W_\rho\ast W_\sigma= \tr(\rho\sigma)\geq 0
\end{equation}
the $P$ function will be $\geq0$ if $g(2\eta ^T \mu_t \eta-C_{1/4})$ is a Wigner function. In 2 dimensions this is true if, and only if, (Ref\cite{simon})

\begin{equation}
\det(2\eta ^T \mu_t \eta - C_{1/4})>\frac{1}{4}
\end{equation}
This positivity of the $P$ function manifests the density matrix as a probability distribution over coherent state projectors, which is therefore of the form (\ref{dis}).

For long times (i.e. once the exponentials and constants are negligible), we have

\begin{equation}
2\eta ^T \mu_t \eta-C_{1/4}\sim 2\left( \begin{array}{cc}
\left(D_{pp}\frac{t}{16\gamma ^3 m^2}+2D_{qp}\frac{t}{4\gamma ^2 m} +D_{qq}t \right) 
& \left(D_{pp}\frac{t}{4\gamma ^2 m}+D_{qq}t\right)\\ \left(D_{pp}\frac{t}{4\gamma ^2 m}+D_{qq}t \right)& D_{pp}t\end{array}\right)
\end{equation}
Due to cancelations, the determinant is

\begin{equation}
2^2\det (D) t^2>0\label{det}
\end{equation}
provided the master equation is Lindblad (by (\ref{D>0})). Thus it is clear that we have complete disentanglement in finite time.
%(I'm assuming the $\det$ condition
%is the hard one to satisfy rather than the $\tr$. It's certainly the harder to check.) 
Exact determination
of the timescale requires numerical methods.

\subsubsection{The limit of no dissipation: $\gamma \to 0$}

The limit $\gamma\to0$ allows comparison with Di\'{o}si's calculation. One finds

\begin{equation}
2\eta ^T \mu_t \eta\to 2Dt + 2\left( \begin{array}{cc}\frac{D_{pp}}{3m^2}t^3 + \frac{D_{qp}}{m}t^2
& \frac{D_{pp}}{2m}t^2\\\frac{D_{pp}}{2m}t^2 & 0\end{array}\right)
\end{equation}
which, allowing for differences in notation, is the same as in Ref\cite{DK} and gives (for a master equation of the form (\ref{Devol}), i.e.  for $D_{qq}=D_{qp}=0$) the disentanglement timescale as calculated there: 

\begin{equation}
t_*\approx1.97\sqrt{\frac{m}{2D_{pp}}}\label{Kiefer}
\end{equation}

\subsection{A particular choice of Lindblad operator}

We consider the `minimally invasive' modification that results in an evolution equation of Lindblad form, and which satisfies
the fluctuation-dissipation theorem (see e.g.Ref\cite{breuer}).   This means we take the Lindblad operator to be

\begin{equation}
L=\sqrt{4mkT}x +i\gamma\sqrt{\frac{1}{4mkT}}p
\end{equation}
resulting in a master equation with

\begin{eqnarray}
D_{pq}&=&0\nonumber\\
D_{pp}&=&2mkT\gamma \nonumber\\
D_{qq}&=&\frac{\gamma}{8mkT}
\end{eqnarray}
This equation has been used in the study of realistic models, such as  the quantum brownian motion model in certain regimes (e.g. Ref\cite{zoup}).
Di\'{o}si's criterion

\begin{equation}\
\det (2\eta ^T \mu_t \eta -C_{1/4})\geq \frac{1}{4}
\end{equation}
becomes

\begin{equation}
\det\left( \begin{array}{cc}
\left(2D_{pp} \frac{4\gamma t + 4e^{-2\gamma t}-e^{-4\gamma t}-3}
{16\gamma ^3 m^2}+2D_{qq}t-\frac{1}{\sqrt{2D_{pp}m}}\right)
& \left(D_{pp}\frac{2\gamma t + e^{-2\gamma t}-1}{2\gamma ^2 m}+D_{qq}t-\frac{1}{2}\right)
\\  \left(D_{pp}\frac{2\gamma t + e^{-2\gamma t}-1}{2\gamma ^2 m}+2D_{qq}t-\frac{1}{2} \right)& 2D_{pp}t -\sqrt{\frac{D_{pp}m}{2}}\end{array}\right)
\geq \frac{1}{4}
\end{equation}
reducing to

\begin{eqnarray}
\Phi \times\left( (\frac{kT}{\gamma})^2(\gamma t)-\frac{1}{4}(\frac{kT}{\gamma})^{3/2}\right) + (\gamma t)^2
-\left(\frac{1}{4}\sqrt{\frac{\gamma}{kT}}+2\sqrt{\frac{kT}{\gamma}}\right)(\gamma t)\nonumber\\
-(\frac{kT}{\gamma})^2\Theta^2
+(\frac{kT}{\gamma})\Theta\geq 0
\end{eqnarray}
or, better, 

\begin{eqnarray}
\frac{\Phi}{4(\gamma t_0)^3}(4\tau -1)-(2+\frac{1}{4}(\gamma t_0)^2)\tau +(\gamma t_0)^2\tau^2\nonumber\\
-\frac{\Theta^2}{(\gamma t_0)^4}+\frac{\Theta}{(\gamma t_0)^2}\geq 0
\end{eqnarray}
with

\begin{equation}
\tau=\frac{t}{t_0}
\end{equation}
and 

\begin{eqnarray}
\Phi&=&4(\gamma t_0)\tau +4e^{-2\gamma t_0 \tau}-e^{-\gamma t_0 \tau}-3\nonumber\\
\Theta&=&2(\gamma t_0)\tau +e^{-2\gamma t_0 \tau}-1
\end{eqnarray}
and where we have introduced a decoherence timescale, $t_0$, given by

\begin{equation}
t_0=\frac{1}{\sqrt{\gamma kT}}
\end{equation}
The decoherence timescale - the time taken for reductions of $e^{-1}$ in quantities measuring the progress of decoherence - is significant because, for reasons discussed in Ref\cite{us} and as found in practice, disentanglement and decoherence occur on comparable timescales. Decoherence is generally more familiar and has been better studied than entanglement breaking (however see Ref\cite{eisert}, Ref\cite{raj}), and so it is of interest to compare the two timescales by using the ratio $\tau$.

If we are at `high' temperatures, i.e. with respect to the damping (or put differently, if decoherence occurs much more quickly than damping, as is the case for situations of emergent classicality): 

\begin{equation}
\gamma t_0=\sqrt{\frac{\gamma}{kT}}\ll1
\end{equation}
we can conjecture a solution to the above equation in the form of a power series in $(\gamma t_0)$. To second order we have

\begin{eqnarray}
\Phi&=&(\gamma t_0)(-3\tau) +(\gamma t_0)^2(\tau^2)\\
\Theta&=&(\gamma t_0)^2(2\tau^2)
\end{eqnarray}
which yields the complete disentanglement time $\tau_*$(as a fraction of the decoherence time)

\begin{equation}
\tau_* = \frac{1}{4} -\frac{25}{48}(\gamma t_0)^2
\end{equation}
This then represents the timescale for this system to become completely disentangled, i.e. disentangled from all other isolated systems.This particular form shows that damping does not affect the overall behaviour, and in this regime shortens the timescale for its occurrence. The result is not directly comparable with (\ref{Kiefer}) because  of the distinct choice of $D$.

If the damping is extremely strong, a better result may be obtained by using (\ref{det}) directly to estimate the timescale.

\section{Disentanglement of Coupled systems}

In this section we consider the separation of two harmonically coupled particles, each in interaction with an environment. Without the coupling the particles might evolve according to (say)

\begin{equation}
\frac{\partial W}{\partial t}=-\frac{p_1}{m_1}\frac{\partial
W}{\partial q_1}  + D_{pp}\frac{\partial ^2
W}{\partial p_1^2}-\frac{p_2}{m_2}\frac{\partial
W}{\partial q_2}  + D_{pp}\frac{\partial ^2
W}{\partial p_2^2}\label{uncoup}
\end{equation}
The `reduced' Wigner function for particle $1$ would then evolve according to exactly (\ref{Devol}). By the disentanglement results above, the particle-$1$ Wigner function would have to be completely disentangled after a finite time. In particular, this means that the full Wigner function for particles $1$ and $2$ separates. As observed above, the results on complete disentanglement specialise to results on $N$-partite separation, unless the particles are coupled in some way.

In this section therefore, we consider the case of two particles which are allowed to `talk' to each other via a harmonic potential. (More general potentials being considerably more difficult to treat.) A potential can act to build up correlations between each party and can countermand some of the disentangling effects of environmental noise.

Once we allow the particles oscillating modes, we must abandon all hope of simple analytic expressions estimating disentanglement times. Graphs of quantities describing the level of entanglement  now will have `wobbles' in them, in addition to any simple overall trends. The precise disentanglement times will depend in detail on the relationship between parameter values, which determine the size and phase of the wobbles. What is more, there is no reason to think that the coupling could not allow the pair of particles to resurrect their entanglement  (in fact, see Ref\cite{raj}). We therefore content ourselves with determining conditions on the couplings that guarantee disentanglement between the pair for good, after a finite time; without worrying about the  precise details and timings of the separation.

\subsection{The Evolution}
We will use the `capital' , rotated coordinate system given by

 \begin{eqnarray}
\bar{Q}_1 &=& \frac{1}{\sqrt{2}}(\bar{q}_1 + \bar{q}_2)\nonumber\\
\bar{Q}_2 &=& \frac{1}{\sqrt{2}}(\bar{q}_1 - \bar{q}_2)\nonumber\\
\bar{P}_1 &=& \frac{1}{\sqrt{2}}(\bar{p}_1 + \bar{p}_2)\nonumber\\
\bar{P}_1 &=& \frac{1}{\sqrt{2}}(\bar{p}_1 - \bar{p}_2)
\end{eqnarray}
or

\begin{equation}
\bar{Z}= \frac{1}{\sqrt{2}}\left( \begin{array}{cc}1_2 & 1_2\\
1_2 & -1_2\end{array}\right)\bar{z}:=R\bar{z}
\end{equation}
The differential equations relevant to the method of characteristics are

\begin{eqnarray}
\frac{d\bar{Q}_1}{dt}&=&-\frac{\bar{P}_1}{m}-2\gamma \bar{Q}_1\nonumber\\
\frac{d\bar{Q}_2}{dt}&=&-\frac{\bar{P}_2}{m}-2\gamma \bar{Q}_2\nonumber\\
\frac{d\bar{P}_1}{dt}&=& m\Omega ^2 \bar{Q}_1\nonumber\\
\frac{d\bar{P}_2}{dt}&=&(2m\omega ^2 + m\Omega ^2)\bar{Q}_2:=m{\Omega'}^2 \bar{Q}_2
\end{eqnarray}
where $\omega$ is the frequency of the inter-particle coupling and $\Omega$ the strength of the well.

\emph{The 1 and 2 evolutions then  decouple and take the same form:}

\begin{equation}
\frac{d}{dt}\bar{Z}=\left( \begin{array}{cc}-2\gamma & -m^{-1}\\
m\omega^2 & 0\end{array}\right)\bar{Z}
\end{equation}
This decoupling will allow us to pursue most of the calculations in the `dashed' and `undashed' sectors in $2$ dimensions, and in parallel. Where we use the same notation at both the $2\times 2$ and $4\times 4$ level, we intend it to be understood that the $4\times 4$ matrix is block diagonal and made up from the dashed and undashed matrices of the same name.

Solving this differential equation we get

\begin{equation}
E_t=\left( \begin{array}{cc}e^{-\gamma t}(\cos \alpha t + \frac{\gamma}{\alpha}\sin\alpha t) &
 -e^{-\gamma t}\frac{\sin \alpha t}{m\alpha}\\
m\Omega^2e^{-\gamma t}\frac{\sin \alpha t}{\alpha} 
& e^{-\gamma t}(\cos \alpha t + \frac{\gamma}{\alpha}\sin\alpha t)\end{array}\right)
\end{equation}
where $\alpha=\sqrt{\Omega^2-\gamma^2}$

The diffusion matrix remains  the same in the rotated coordinates. We will take the diffusion coefficient matrix to be diagonal, having in mind eventually to use the minimal Lindblad values as above.  We obtain

\begin{eqnarray}
\mu _t &=&\int_0 ^t d\tau E_{\tau} ^T \left( \begin{array}{cc}D_1 & 0\\
0 & D_2\end{array}\right) E_{\tau}\nonumber\\
&=&\int_0^t d\tau e^{-2\gamma t}\left( \begin{array}{cc}a&  b\\b&c\end{array}\right)\nonumber\\
&=&\left( \begin{array}{cc}A &  B\\B&C\end{array}\right)
\end{eqnarray}
where

\begin{eqnarray}
a &=& \frac{D_1}{2}\left(1+\frac{\gamma^2}{\alpha^2}\right)
+\frac{D_2m^2\Omega^4}{2\alpha^2}
+\frac{D_1\gamma}{\alpha}\sin 2\alpha\tau+
\cos 2\alpha\tau
\left(\frac{D_1}{2}(1-\frac{\gamma^2}{\alpha^2})
-\frac{D_2m^2\Omega^4}{2\alpha^2}\right)\nonumber\\
b&=&\left(\frac{D_1\gamma}{2m\alpha^2}-\frac{D_2m\Omega^2\gamma}{2\alpha^2}\right)
\left(-1-\frac{\alpha}{\gamma}\sin 2\alpha\tau+\cos 2\alpha\tau\right)\nonumber\\
c&=&\frac{D_1}{2m^2\alpha^2}+ \frac{D_2}{2}\left(1+\frac{\gamma^2}{\alpha^2}\right)+\frac{D_2\gamma}{\alpha}\sin 2\alpha\tau
 +\cos 2\alpha\tau
 \left(\frac{D_2}{2}(1-\frac{\gamma^2}{\alpha^2})-\frac{D_1}{2m^2\alpha^2}\right)\nonumber\\
\end{eqnarray}
and

\begin{eqnarray}
A&=&\left(\frac{D_1}{2}(1+\frac{\gamma^2}{\alpha^2})+\frac{D_2m^2\Omega^4}{2\alpha^2}\right)\frac{1-e^{-2\gamma t}}{2\gamma}+
\frac{D_1\gamma}{2\alpha(\gamma^2+\alpha^2)}(\alpha-\alpha e^{-2\gamma t}\cos 2\alpha t
-\gamma e^{-2\gamma t}\sin 2\alpha t)\nonumber\\
&&+\left(\frac{D_1}{2}(1-\frac{\gamma^2}{\alpha^2})-\frac{D_2m^2\Omega^4}{2\alpha^2}\right)
\frac{\gamma -\gamma e^{-2\gamma t}\cos 2\alpha t+\alpha e^{-2\gamma t}\sin 2\alpha t}{2(\gamma^2+\alpha^2)}\nonumber\\
B&=&\left(\frac{D_1\gamma}{2m\alpha^2}-\frac{D_2m\Omega^2\gamma}{2\alpha^2}\right)
\left(\frac{e^{-2\gamma t}-1}{2\gamma}
-\frac{\alpha(\alpha-\alpha e^{-2\gamma t}\cos 2\alpha t-\gamma e^{-2\gamma t}\sin 2\alpha t)}{2\gamma(\gamma^2+\alpha^2)}\right)\nonumber\\
C&=&\left(\frac{D_1}{2m^2\alpha^2}+ \frac{D_2}{2}(1+\frac{\gamma^2}{\alpha^2})\right)\frac{1-e^{-2\gamma t}}{2\gamma}+
\frac{D_2\gamma}{2\alpha(\gamma^2+\alpha^2)}(\alpha-\alpha e^{-2\gamma t}\cos 2\alpha t
-\gamma e^{-2\gamma t}\sin 2\alpha t)\nonumber\\
&&+\left(\frac{D_2}{2}(1-\frac{\gamma^2}{\alpha^2})-\frac{D_1}{2m^2\alpha^2}\right)
\frac{\gamma -\gamma e^{-2\gamma t}\cos 2\alpha t+\alpha e^{-2\gamma t}\sin 2\alpha t}{2(\gamma^2+\alpha^2)}
\end{eqnarray}
With all this, we have

\begin{equation}
W_t(Z)=g_{M_t}\ast W_0(\epsilon_t Z)|\det E_t|
\end{equation}
where

\begin{equation}
\epsilon_t=\eta^T E_t^T\eta
\end{equation}
and

\begin{equation}
M_t=2\eta^T \mu_t \eta
\end{equation}

\subsection{Showing separation}
Now we have solved for the evolution, how are we to address the issue of separation? Simple, general criteria for separability of states are notoriously difficult to come by. There are three notable exceptions: $\mathcal{H}=\mathbb{C}^2\otimes\mathbb{C}^2$; $\mathcal{H}=\mathbb{C}^2\otimes\mathbb{C}^3$; and, embedded within $\mathcal{H}=L^2(\mathbb{R})\otimes L^2(\mathbb{R})$, the class of gaussian states. Separation criteria for gaussian states come in a variety of flavours (see Ref\cite{gauss} and for relations between them Ref\cite{Giedke}); we shall use the particularly simple one from the paper of Duan \emph{et al.}. In terms of our rotated coordinate system, this criterion reads: a Wigner function, $g(\Sigma)$, is separable if, and only if

\begin{eqnarray}
\Sigma_{11}+\Sigma_{44}\geq 1\nonumber\\
\Sigma_{22}+\Sigma_{33}\geq 1\label{duan}
\end{eqnarray}
Separation criteria for gaussian states will be useful to us  because the propagators for our evolution are gaussian. In words: if , after a certain time, we can take off part of the propagator and use it to smear the initial Wigner function positive, and still be left over with a gaussian that separates,  our evolution will be a separating one. In symbols: 

\begin{eqnarray}
\frac{W_t(Z)}{|\det E_t|}&=&g_{M_t}\ast W_0(\epsilon_t Z)\nonumber\\
&=:&\epsilon_t^*(g_{M_t}\ast W_0)(Z)\nonumber\\
&=&(g_{\epsilon_{-t} M_t \epsilon_{-t}^T}\ast \epsilon_t^*W_0)(Z)\nonumber\\
&=&(g_{\epsilon_{-t} M_t \epsilon_{-t}^T -C}\ast g_C\ast\epsilon_t^*W_0)(Z)
\end{eqnarray}
which means we can examine to see whether

\begin{enumerate}
\item the Gaussian $g_{\epsilon_{-t} M_t \epsilon_{-t}^T -C}$ separates
\item we can simultaneously choose $C$ such that $g_C\ast\epsilon_t^*W_0 \geq 0$.
\end{enumerate}

If we can do both these things, then $W_t(z)$ separates. It will be of the form

\begin{equation}
\sum_a p_a \int d^2 z_1' d^2 z_2'p(z_1',z_2')W_a^{(1)}(z_1 -z_1')W_a^{(2)}(z_2-z_2')
\end{equation}
where

\begin{equation}
p=g_C\ast\epsilon_t^*W_0|\det E_t|
\end{equation}
and the (not necessarily discrete) sum over $a$ describes the separation of the state
$g_{\epsilon_{-t} M_t \epsilon_{-t}^T -C}$.

\subsection{Calculations relevant to separation}

Pursuing our second point, it is straightforward to see that

\begin{equation}
g_C*\epsilon_t^*W_0=\epsilon_t^*(g_{\epsilon_t C\epsilon_t^T}*W_0)
\end{equation}
so that this will be $\geq 0$ if $g_{\epsilon_t C\epsilon_t^T}$ is a Wigner function.
 When is $g_{\epsilon_tC\epsilon_t^T}$  a Wigner function ?
 The `capital' or
rotated coordinate system will be used unless otherwise stated. If we have a gaussian $g_M$ with

\begin{equation}
M=\left( \begin{array}{cc} m & 0\\
0 & m'\end{array}\right)
\end{equation}
and the matrices $m$ and $m'$ give rise to $2\times2$ gaussians each of which is a Wigner function, then $g_M$ is Wigner.
We know that this is the case if

\begin{eqnarray}
 m+\frac{i}{2}\eta&\geq&0\nonumber\\
 m'+\frac{i}{2}\eta&\geq&0
\end{eqnarray}
which in 2 dimensions is equivalent to

\begin{eqnarray}
\det m &\geq& \frac{1}{4}\nonumber\\
\det m' &\geq& \frac{1}{4}
\end{eqnarray}
as long as the traces are positive.

In our chosen coordinate system, the relevant matrices may be taken to be block diagonal. The majority of the calculations then
follow along in $2$ dimensions. We pick $C$ block diagonal and such that

\begin{equation}
\epsilon_t C\epsilon_t^T=C_{0}
\end{equation}
with $C_0$ positive enough to give a Wigner function, i.e.

\begin{equation}
\det C_0 =\frac{1}{4}
\end{equation}
in both the dashed and undashed sectors. Thus $g_{\epsilon_t C\epsilon_t^T}$ is a wigner function \emph{by construction}.

Consistent with this, we must now address our first point. We want $g_{\epsilon_{-t} M_t \epsilon_{-t}^T -C}$ to be a separable Wigner
function. First, we must check it is a Wigner function. This will turn out to be the difficult condition to satisfy, in the sense that the separation criterion, while messy to check, follows on trivially if  $g_{\epsilon_{-t} M_t \epsilon_{-t}^T -C}$ does become a Wigner function. We work a the level of the $2\times 2$ blocks again

We want

\begin{eqnarray}
&&\det (\epsilon_{-t} M_t \epsilon_{-t}^T -C)\nonumber\\
&=& \det E_{-t}^2\det (2\mu_t - (\eta C_{0}\eta^T))\nonumber\\
&=&e^{4\gamma t}\det (2\mu_t - (\eta C_{0}\eta^T))\geq \frac{1}{4}
\end{eqnarray}
We choose

\begin{equation}
\eta C_{0}\eta^T=\left( \begin{array}{cc}m\sqrt{\gamma kT} & 1/2\\
1/2 & 1/2m\sqrt{\gamma kT}\end{array}\right)
\end{equation}
and for reasons explained above, declare ourselves only interested in the asymptotics, and conditions guaranteeing finite-time termination of entanglement. Therefore we throw away
any term containing $e^{-2\gamma t}$. We also proceed with $D_1$ and $D_2$ chosen as in the `minimal' Lindblad case above.
This means

\begin{eqnarray}
A&\sim&\frac{D_1(\Omega^4+3\gamma^2\Omega^2-4\gamma^4)}{4\gamma\Omega^2(\Omega^2-\gamma^2)}+\frac{D_2m^2\Omega^2}{4\gamma}+O(e^{-2\gamma t})\nonumber\\
&=&\frac{mkT(\Omega^4+3\gamma^2\Omega^2-4\gamma^4)}{2\Omega^2(\Omega^2-\gamma^2)}+\frac{m\Omega^2}{32kT}+O(e^{-2\gamma t})\nonumber\\
B&\sim&\frac{m^2\Omega^2D_2-D_1}{2m\Omega^2}+O(e^{-2\gamma t})\nonumber\\
&=&\frac{\gamma}{16kT}-\frac{kT\gamma}{\Omega^2}+O(e^{-2\gamma t})\nonumber\\
C&\sim&\frac{D_1}{4m^2\gamma\Omega^2}+\frac{D_2(\Omega^4+3\gamma^2\Omega^2-4\gamma^4)}{4\gamma\Omega^2(\Omega^2-\gamma^2)}+O(e^{-2\gamma t})\nonumber\\
&=&\frac{kT}{2m\Omega^2}+\frac{(\Omega^4+3\gamma^2\Omega^2-4\gamma^4)}{32mkT\Omega^2(\Omega^2-\gamma^2)}+O(e^{-2\gamma t})
\end{eqnarray}
So, ignoring terms $O(e^{-2\gamma t})$,

\begin{eqnarray}
\det(2\mu_t - (\eta C_{0}\eta^T))&=&\frac{(\Omega^4+3\gamma^2\Omega^2-4\gamma^4)}{(\Omega^2-\gamma^2)\Omega^2}
\left(\frac{(kT)^2}{\Omega^2}-1/2\sqrt{\frac{kT}{\gamma}}+\frac{\Omega^2}{(16kT)^2}-1/16\sqrt{\frac{\gamma}{kT}}\right)\nonumber\\
&&+5/16 +\frac{(\Omega^4+3\gamma^2\Omega^2-4\gamma^4)^2}{16\Omega^4(\Omega^2-\gamma^2)^2}-\frac{\Omega^2}{32\sqrt{\gamma (kT)^2}}
-\frac{\sqrt{\gamma (kT)^2}}{\Omega^2}-4(\frac{kT\gamma}{\Omega^2})^2\nonumber\\
&&-(\frac{\gamma}{8kT})^2+\frac{\gamma^2}{2\Omega^2}-\frac{2kT\gamma}{\Omega^2}+\frac{\gamma}{8kT}\label{crit}
\end{eqnarray}
If this last quantity (consisting of various dimensionless comparisons between the constants present) is positive definite, our Wigner
condition will be satisfied on a timescale set by $\gamma$. It is not transparent which regimes this will be positive
in, but at high temperatures, the important term is 

\begin{equation}
\left(\frac{kT\gamma}{\Omega^2}\right)^2\left(\frac{(\Omega^4+3\gamma^2\Omega^2-4\gamma^4)}{\gamma^2(\Omega^2-\gamma^2)}-4\right)
=(t_{osc}/t_{dec})^4\left(\frac{(\Omega^4+3\gamma^2\Omega^2-4\gamma^4)}{\gamma^2(\Omega^2-\gamma^2)}-4\right)
\end{equation}
with $t_{dec}=t_0$ and $t_{osc}=\Omega^{-1}$. This will be positive if the system is under-damped, i.e. $\Omega^2>\gamma^2$.

\emph{Note:} if the particles are both sitting in a harmonic well, this condition will need to be satisfied for both $\Omega$ and $\Omega'$.

We must now check that separability will also occur. Above,(\ref{duan}), we stated the Duan \emph{et al.} criteria for separability.
Note that one could equally use $\eta \Sigma \eta^T$ instead of $\Sigma$. Applying this to $g_{\epsilon_{-t} M_t \epsilon_{-t}^T -C}$, we become
interested in

\begin{eqnarray}
\eta \epsilon_{-t}(M_t - C_{0})\epsilon_{-t}^T \eta^T
=E_{-t}^T(2\mu_t - (\eta C_{0}\eta^T))E_{-t}
\end{eqnarray}
Now

\begin{eqnarray}
&&\langle Q_{1,2}^2+P_{2,1}^2\rangle_{g_{E_{-t}^T(2\mu_t - (\eta C_{0}\eta^T))E_{-t}}}=\nonumber\\
&&\int dZ (Q_{1,2}^2+P_{2,1}^2)\frac{1}{\pi^2\sqrt{\det 2E_{-t}^T(2\mu_t - (\eta C_{0}\eta^T))E_{-t}}}\exp
\left[-Z^T \frac{1}{2E_{-t}^T(2\mu_t - (\eta C_{0}\eta^T))E_{-t}}Z\right]\nonumber\\
&=&\int dX (E_{-t}^TX)^TM_{1,2}(E_{-t}^TX)g_{2\mu_t - (\eta C_{0}\eta^T)}(X)\nonumber\\
&=&\langle (E_{-t}^TX)^Tm_{1,2}(E_{-t}^TX)\rangle_{g_{2\mu_t - (\eta C_{0}\eta^T)}^{(2\times 2)}}
+ \langle({E_{-t}^T}'X')^Tm_{2,1}({E_{-t}^T}'X') \rangle_{g_{2\mu_t' - (\eta C_{0}'\eta^T)}^{(2\times 2)}}\nonumber\\
&=&\tr (E_{-t}^Tm_{1,2}E_{-t}\langle XX^T\rangle_{g_{2\mu_t - (\eta C_{0}\eta^T)}^{(2\times 2)}})+
\tr ({E_{-t}^T}'m_{1,2}{E_{-t}}'\langle X'{X^T}'\rangle_{g_{2\mu_t' - (\eta C_{0}'\eta^T)}^{(2\times 2)}})\nonumber\\
&=&\tr (E_{-t}^Tm_{1,2}E_{-t}(2\mu_t - (\eta C_{0}\eta^T)))+\tr ({E_{-t}^T}'m_{1,2}E_{-t}'(2\mu_t' - (\eta C_{0}'\eta^T)))\nonumber\\
&=:& (1,2)+(1,2)'
\end{eqnarray}
where

\begin{eqnarray}
M_{1,2}&=&\left( \begin{array}{cc}m_{1,2} & 0\\
0 & m_{2,1}\end{array}\right)\nonumber\\
m_1&=&\left( \begin{array}{cc}1 & 0\\
0 & 0\end{array}\right)\nonumber\\
m_2&=&\left( \begin{array}{cc}0 & 0\\
0 & 1\end{array}\right)\nonumber\\
X^T&=:&(X^T,X'^T)
\end{eqnarray}
and everything is still block diagonal.

\begin{equation}
E_{-t}^Tm_1E_{-t}=e^{2\gamma t}\left( \begin{array}{cc}(\cos \alpha t -\frac{\gamma}{\alpha}\sin \alpha t)^2
& \frac{\sin \alpha t}{m \alpha}(\cos \alpha t-\frac{\gamma}{\alpha}\sin \alpha t)\\
\frac{\sin \alpha t}{m \alpha}(\cos \alpha t-\frac{\gamma}{\alpha}\sin \alpha t)
 & ( \frac{\sin \alpha t}{m\alpha} ) ^2 \end{array}\right)
\end{equation}
and

\begin{equation}
E_{-t}^Tm_2E_{-t}=e^{2\gamma t}\left( \begin{array}{cc}(\frac{m\Omega^2\sin \alpha t}{\alpha})^2
& -\frac{m\Omega^2\sin \alpha t}{ \alpha}(\cos \alpha t-\frac{\gamma}{\alpha}\sin \alpha t)\\
-\frac{m\Omega^2\sin \alpha t}{ \alpha}(\cos \alpha t-\frac{\gamma}{\alpha}\sin \alpha t)
 & (\cos \alpha t -\frac{\gamma}{\alpha}\sin \alpha t)^2 \end{array}\right)
\end{equation}
In fact

\begin{equation}
\tr(E_{-t}^Tm_{1,2}E_{-t}N)=v_{1,2}^TNv_{1,2}
\end{equation}
where $v_{1,2}$ are the vectors given by the first and second columns of $E_{-t}$ read upwards. This means that

\begin{equation}
\tr(E_{-t}^Tm_{1}E_{-t}(2\mu_t - (\eta C_{0}\eta^T)))=e^{2\gamma t}u^T\left( \begin{array}{cc}1 & 0\\
-\gamma/\alpha& 1/m\alpha\end{array}\right)(2\mu_t - (\eta C_{0}\eta^T))\left( \begin{array}{cc}1 & -\gamma/\alpha\\
0 & 1/m\alpha\end{array}\right)u
\end{equation}
and

\begin{equation}
\tr(E_{-t}^Tm_{2}E_{-t}(2\mu_t - (\eta C_{0}\eta^T)))=e^{2\gamma t}u^T\left( \begin{array}{cc}0 & 1\\
-m\Omega^2/\alpha & -\gamma/\alpha\end{array}\right)(2\mu_t - (\eta C_{0}\eta^T))
\left( \begin{array}{cc}0 & -m\Omega^2/\alpha\\
1 &-\gamma/\alpha\end{array}\right)u
\end{equation}
where $u^T=(\cos \alpha t,\sin\alpha t)$, a euclidean unit vector. This is still a mess, but this form allows us to see that, as long as we satisfy $(\ref{crit})>0$ (the Wigner condition), the separability condition will come to be satisfied on a timescale set most importantly by $\gamma$. 

\subsection{The case of $\gamma=0$}

We now consider the case of $\gamma=0$. For $\gamma=0$ and a diagonal diffusion matrix, we can proceed as above with much simplified equations. We find

\begin{eqnarray}
A&=&\left(\frac{D_{pp}}{2} + \frac{D_{qq}m^2\Omega^2}{2}\right)t
+\left(\frac{D_{pp}}{2} - \frac{D_{qq}m^2\Omega^2}{2}\right)\frac{\sin\Omega t}{2\Omega}\nonumber\\
B&=&0\nonumber\\
C&=&\left(\frac{D_{pp}}{2m^2\Omega^2} + \frac{D_{qq}}{2}\right)t + \left(\frac{D_{qq}}{2} - \frac{D_{pp}}{2m^2\Omega^2}\right)\frac{\sin\Omega t}{2\Omega}
\end{eqnarray}
The requisite conditions will obviously come to be satisfied after a finite time, guaranteed by $D>0$: the propagator spreads out in a linear way and the evolution is

\begin{equation}
E_t=\left( \begin{array}{cc}\cos \Omega t  & -\frac{\sin \Omega t}{m\Omega}\\
m\Omega\sin \Omega t
& \cos \alpha t \end{array}\right)
\end{equation} 
which clearly doesn't stop $E_{-t}^T\mu_tE_t$ from coming to satisfy the separability criteria.

\section{Discussion}

Given the similarity between the results in Section 2 and Section 3 it is probably worth summarising and discussing in what ways they are related. In Section 2, we considered the evolution of a system according to  (\ref{evol}). We showed that after a finite time, the map effecting the evolution came into the form

\begin{equation}
M(\rho)=\sum_k\tr(\mu_k\rho)m_k\label{dis2}
\end{equation}
which meant that system had become disentangled from all other systems.  What happens if the system of interest is itself composite? If we imagine that the $\rho$ in (\ref{dis2}) is actually a state on $A\otimes B$, then we could enquire about the form of the reduced state on $A$ after such an evolution. Let us expand the initial state $\rho$ as

\begin{equation}
\rho=\sum_{ij}\pi_{ij}\rho_i^A\otimes\rho_j^B
\end{equation}
where the $\{\pi_{ij}\}$ are not necessarily positive. Then we have 

\begin{eqnarray}
\tr_B M(\rho)&=&\sum_k \tr_A\left(\tr_B\left(\mu_k\rho\right) \right)\tr_Bm_k\nonumber\\
&=&\sum_{kij}\tr_A\left(\rho_i^A\pi_{ij}\left(\tr_B\mu_k\rho_j^B\right)\right)\tr_Bm_k
\end{eqnarray}
The map governing the evolution of $\tr_B\rho$ will therefore not usually be of the completely disentangling form unless $\rho$ is factored in the first place. 

An exception to this is if the systems $A$ and $B$ do not interact, in the sense that their evolution map takes the factored form $M=M_A\otimes M_B$. This is the case for instance, when the coupling $\omega$ of Section 3 vanishes, as in (\ref{uncoup}). When this happens there are closed equations governing the evolution of each subsystem, and they each result in a completely disentangling evolution. In particular  then, systems $A$ and $B$ must  disentangle. Seen in this context, the point of Section 3 was to investigate how much coupling between two systems spoils their disentanglement by a noisy environment. The answer to the question is essentially that it does not, at least for the conditions of interest for emergent classicality.

 Of course, Section 3 also illustrated how the technology devised to test for separability in gaussian states can be used to investigate dynamical disentanglement  in bipartite systems for quite general states. 

All this has further confirmed expectations that uncontrolled environmental interactions are highly destructive to entanglement in quantum systems. That this destruction occurs (and occurs quickly) is not surprising given our experience of decoherence in such systems, and since entanglement relies on the superposition principle for its existence. It is interesting though, that by contrast with decoherence, entanglement - even with \emph{all} isolated systems - can be wiped out totally after a finite time. 
When the entangled systems are allowed to interact with each other, the situation is less simple as the interaction can help to rebuild lost entanglement. However, in certain regimes (i.e. for certain regions in the space of temperature, damping coefficients e.t.c.), the entanglement between the systems will still be totally destroyed after a finite time.

Although it may be of practical interest to know about the effects of environmental interactions on entangled systems (so as to avoid it in quantum technologies), in investigating these questions we have been primarily interested in emergent classical behaviour. Entanglement \emph{is} an inherently quantum phenomenon, and so it is encouraging to find an efficient mechanism by which it is rapidly destroyed in the systems of our everyday experience. But to what extent,  and in what sense, is the absence of entanglement necessary for a system to admit a classical description? Entanglement between two systems $A$ and $B$ certainly means that we cannot hope to describe their correlations in a classical way. For example, if they remain entangled under evolution, we would not expect to find decoherent histories which refer to each system separately. In the situations where we normally try to explain the existence of a classical description however, there is not normally this clean division into  definite subsystems. Given a gas of particles in a box, what sort of entanglement properties do we need to destroy in order to be able to describe it classically? There is no quantitative answer to this as yet. But a picture hinted at here is that we would like the subsystems which we refer to by the coarse-grained observables of classical interest to disentangle. For instance, suppose we were interested in local number, energy and momentum densities: $n(\mathbf{x})$, $h(\mathbf{x})$ and $p(\mathbf{x})$(see e.g. Ref\cite{hall}).  Near equilibrium, we expect to be able to describe the system by states which are approximate eigenvectors of $n$, $h$ and $p$. What we mean is that there should be no entanglement with respect to these indices in order for us to be able to find a classical description.

\section{Acknowledgements}
The author would like to thank J.J.Halliwell and L.Di\'{o}si for some very useful discussions and both PPARC and the British Council for their financial support.

\end{document}